\documentclass[runningheads]{llncs}
\usepackage{graphicx}
\usepackage{epsfig}
\usepackage{calc}
\usepackage{amsmath}
\usepackage{booktabs}
\usepackage{mathpartir}		
\usepackage{xspace}			
\usepackage{enumerate}		
\usepackage{longtable}
\usepackage{color}			
\usepackage{ifthen}			
\usepackage[colorinlistoftodos]{todonotes}
\usepackage{color}

\usepackage{footnote}

\usepackage[normalem]{ulem}
\usepackage{comment}

\usepackage{array}
\newcommand{\PreserveBackslash}[1]{\let\temp=\\#1\let\\=\temp}
\newcolumntype{C}[1]{>{\PreserveBackslash\centering}p{#1}}
\newcolumntype{R}[1]{>{\PreserveBackslash\raggedleft}p{#1}}
\newcolumntype{L}[1]{>{\PreserveBackslash\raggedright}p{#1}}

\usepackage{subcaption}
\usepackage{listings}

\begin{document}

\title{FOSS-chain: using blockchain for Open Source Software license compliance}

\author{Kypros Iacovou\inst{1} \and
Georgia M. Kapitsaki\inst{1}\orcidID{0000-0003-3742-7123} \and
Evangelia Vanezi\inst{1}\orcidID{0000-0003-1958-5574}}

\authorrunning{Iacovou \and Kapitsaki \and Vanezi}

\institute{University of Cyprus, Nicosia, Cyprus \\
\email{kyproslfc7@gmail.com,\{gkapi,vanezi.evangelia\}@ucy.ac.cy}
}

\titlerunning{FOSS-chain: using blockchain for OSS license compliance}

\maketitle              
\begin{abstract}

Open Source Software (OSS) is widely used and carries licenses that indicate the terms under which the software is provided for use, also specifying modification and distribution rules. Ensuring that users are respecting OSS license terms when creating derivative works is a complex process. Compliance issues arising from incompatibilities among licenses may lead to legal disputes. At the same time, the blockchain technology with immutable entries offers a mechanism to provide transparency when it comes to licensing and ensure software changes are recorded. In this work, we are introducing an integration of blockchain and license management when creating derivative works, in order to tackle the issue of OSS license compatibility. We have designed, implemented and performed a preliminary evaluation of \emph{FOSS-chain}, a web platform that uses blockchain and automates the license compliance process, covering 14 OSS licenses. We have evaluated the initial prototype version of the \emph{FOSS-chain} platform via a small scale user study. Our preliminary results are promising, demonstrating the potential of the platform for adaptation on realistic software systems.

\end{abstract}

\section{Introduction}

Open Source Software (OSS) is everywhere, with a large number of OSS repositories being available for reuse online, while there is even participation of commercial companies to OSS development~\cite{qin2025developers}. Open Source Software carries licenses, such as MIT and General Public Licenses (GPL), that regulate the usage, modification and distribution of OSS defining specific licensing terms in their respective legal texts~\cite{rosen2005open}. Nowadays, a vast number of licenses are available for use with the Open Source Initiative (OSI)\footnote{\url{https://opensource.org/}} having approved more than 80 licenses and the Software Package Data Exchange (SPDX) listing more than 550 licenses and exceptions licenses~\cite{stewart2010software}. When creating derivative works, the compatibility between software licenses is important for developers, businesses and the OSS community in general, as it determines whether software components may be legally linked without violating licensing terms. This can create challenges for developers who wish to create derivative works of existing software or combine OSS components with different licensing models, particularly when they intend to release proprietary versions of the software.

At the same time, the blockchain technology can be used to build applications for smart contracts, supply chain management and digital identity verification, among others~\cite{zheng2018blockchain}. Blockchain is designed to offer a decentralized, secure and transparent method for recording transactions. Each transaction gets an entry on the distributed ledger and consensus efforts verify it. This makes it extremely challenging for an adversary to alter or manipulate transactions, as anything added to the blockchain is defined immutable~\cite{kshetri2017blockchain}. Blockchain's security and transparency properties make it suitable for smart contracts involving OSS license tracking and compliance verification, as they are self-executing and do not require human intervention.

Existing works on license compliance focus on source code analysis to understand licensing information or address license compliance as part of Software Composition Analysis (SCA) processes~\cite{ombredanne2020free,tuunanen2009automated}. Such solutions usually require human intervention or provide post-hoc management of licenses. Blockchain can be useful in this respect for proactive management of license compliance, as it can be used to host OSS components and their licensing information, helping developers track software’s usage, modification, and distribution over time. Using a blockchain-based system, allows also software contributors to be recognized automatically as contributors of the original software.

Using the above as starting point, in this work we are using a design science approach and are introducing \emph{\emph{FOSS-chain}}, a platform that integrates blockchain into license compliance management. \emph{FOSS-chain} relies heavily on smart contracts of blockchain, which are immutable agreements that help in managing and checking the licenses when a developer downloads or uploads software projects. Smart contracts are also very useful when a developer relies on existing software in order to create a derivative work. In order to make the system available for use by OSS developers, we have created a web platform where users can create a new account, search for existing software, download it, and share their own software. \emph{FOSS-chain} performs a license compatibility check, when a user uploads a software that is a derivative work of an existing project on-chain, and currently supports 14 popular OSS licenses. By focusing on proactive enforcement rather than post-distribution audits, \emph{FOSS-chain} shifts the compliance process upstream, reducing legal risk and developers' uncertainty on license management. We have performed a preliminary evaluation of the feasibility of \emph{FOSS-chain} and its initial prototype implementation via a small-scale user study, in order to examine its potential. Users of technical and non-technical background have participated in the evaluation. Most users find the platform useful and easy to use, while they have made suggestions for further improvements. Simple statistical analysis has been employed for this part of the work.

The contribution of this work lies in: 1) the introduction of a blockchain-based smart contracts architecture to enforce OSS compliance and preserve licensing records immutably, and 2) the provision of a web platform that integrates blockchain with license management. \textbf{Platform availability.} \emph{FOSS-chain} is available on a GitHub repository online~\cite{replication}. 

The remainder of the text is structured as follows. Section 2 presents background concepts, while related work is described in section 3. Section 4 is dedicated to the presence of the \emph{FOSS-chain} architecture and platform, including its design and prototype implementation. The preliminary evaluation performed is presented in section 5, while section 6 briefly discusses main findings and threats to validity. Finally, section 7 concludes the work.

\section{Background concepts}

\subsection{Open Source Software licensing}

Open Source Software licenses generally fall into three main categories: copyleft and permissive (or non-copyleft) licenses that are further divided into strong and weak copyleft. Strong copyleft licenses, such as the GNU GPL v3.0 (GPL-3.0) and the Affero GPL v3.0 (AGPL-3.0), require that if a work is modified or a derivative incorporating the original software is produced, it must also be released under the same OSS licensing terms. This enables publicly accessible improvements and prevents proprietary versions from being created without sharing modifications. On the other hand, permissive licenses, such as the MIT, the Apache v2.0 (Apache-2.0) and the Berkeley Software Distribution (BSD) licenses (e.g. BSD-2-Clause), have minimal restrictions. They allow developers to modify, use, and distribute the software freely, incorporating it also into proprietary projects. These licenses rely more on community and economic incentives to encourage contributions to the original project. For example, ReactJS carries a MIT License, while TensorFlow is licensed under Apache-2.0. In between, weak copyleft licenses like the GNU Lesser General Public License v2.1 (LGPL-2.1) require the use of the same (or of a compatible) license, when the original software is modified but do not pose these restrictions when the original software remains intact. 

License compatibility refers to having different OSS licenses combined in a software project without legal conflicts. A major difference between permissive and copyleft licenses is that permissive-licensed code can be incorporated into copyleft-licensed projects, but that is not always allowed the other way around, due to the stricter sharing requirements of copyleft licenses~\cite{kapitsaki2015insight}. For instance, permissive licenses, such as MIT or Apache-2.0, allow code to be used in a software system under any license, including proprietary licenses. Software integrating GPL-licensed components must also be made available under GPL (or under a compatible license). These restrictions can create legal and financial risks for companies that incorporate OSS into their products without being fully aware of the licensing implications, and relevant legal disputes can be found in the literature, e.g. Artifex v. Hancom\footnote{\url{https://www.fsf.org/blogs/licensing/update-on-artifex-v-hancom-gnu-gpl-compliance-case-1}} concerning the Ghostscript project.

\subsection{Blockchain}

A blockchain is a distributed ledger, which is a decentralized database that records transactions across a computer network. It was first presented as an underlying mechanism for Bitcoin~\cite{nakamoto2008bitcoin}. Unlike traditional databases that are managed by a central authority, such as banks in a financial network, blockchains operate on the peer-to-peer (P2P) network with no single entity controlling the records. Data are being distributed among all participants in the network.

Immutability is a key feature of blockchain: once a transaction gets added to the blockchain and validated by the network, it cannot get changed or deleted. Transactions are collected into blocks, and each subsequent block is cryptographically bound to the previous one, thus creating a chain. Altering any previous record will necessitate altering all subsequent blocks as well, which cannot happen due to network consensus. A key advantage of blockchain is transparency. Due to the public ledger system of blockchains, all transactions are verifiable by the participants of the network which removes the need for any middlemen and chances of fraud. Smart contracts are self-executing contracts with the terms of the agreement directly written into lines of code. By ensuring that all parties adhere to the predefined conditions without ambiguity, administrative overhead is reduced and the potential for human error is eliminated, minimizing the risk of disputes. In terms of license compliance, smart contracts can be used to improve the enforcement of rules, regulations and contracts.

\section{Related work}

Most prior works on OSS licensing have focused on tools and frameworks that help developers identify license types, detect conflicts, and ensure license compliance. There are works that assist developers to scan the source code of software systems and extract relevant licenses, such as FOSSology~\cite{jaeger2017fossology} that integrates the Ninka license scanner~\cite{german2010sentence} and ASLA~\cite{tuunanen2009automated}. Other works rely on performing SCA for license compliance~\cite{ombredanne2020free}, such as OSSPolice tailored to mobile applications~\cite{duan2017identifying}, or on extracting terms from license texts~\cite{kapitsaki2017identifying}, while recommender systems like findOSSLicense~\cite{kapitsaki2016find,kapitsaki2019modeling} and online resources (e.g. choosealicense,\footnote{\url{https://choosealicense.com/}} TLDRLegal)\footnote{\url{https://www.tldrlegal.com/}} for choosing licenses and detecting incompatibilities (e.g. LiDetector, SPDX compatibility check) also exist~\cite{kapitsaki2015open,xu2023lidetector}. Such tools assist organizations in identifying issues of non-compliance, but operate usually as reactive mechanisms.

Blockchain has been studied in the past in the context of software engineering. An exploratory study on the smart contracts of Ethereum examined all smart contracts that were created via contract creation transactions~\cite{oliva2020exploratory}. It was found that only a very small percentage of the contracts are used in the majority of 
transactions, while high-activity contracts have a very small number of source code instructions. When it comes to blockchain transactions processing time, another work found that properties concerning gas pricing behaviors are highly associated with processing times~\cite{pacheco2023makes}. On the developers side, Rosa et al. examined why and how developers maintain smart contracts using 14 OSS smart contract repositories in Solidity~\cite{rosa2025and}. It was found that developers are mainly making changes to improve the scripts' internal quality and fix bugs.

Existing works in the literature based on blockchain and license compliance are mainly conceptual, or proof-of-concept-based. Blockchain has been suggested as a measure to control software piracy, with smart contracts being used to enforce licensing agreements~\cite{shamalka2024blockchain}. In its initial development, this prior work has considered a single software offered by a software vendor who is the owner of the platform and does not consider OSS. The proposed tool does not seem to be available online. Another prior work has introduced the concept of using blockchain for OSS license management similarly as in the current work~\cite{kumar2022approach}. The authors used InterPlanetary File System (IPFS), smart contracts, transaction manager (Meta-Mask) and a permissioned blockchain (based on the Ethereum platform) to enforce the conformance of licenses, whereas they also covered commercialization of a software project. Although this later work is very close to our work, it offers a simulation of the concept and does not consider specific OSS licenses and relevant compatibilities. Moreover, in our prototype version of \emph{FOSS-chain} we are providing some additional features to users interacting with the platform (e.g. project search). Overall, the above prior studies focus more on license recording and visibility. 

\section{The FOSS-chain platform}

\subsection{Platform overview and blockchain use}

\begin{figure}
\centering
\includegraphics[scale=0.38]{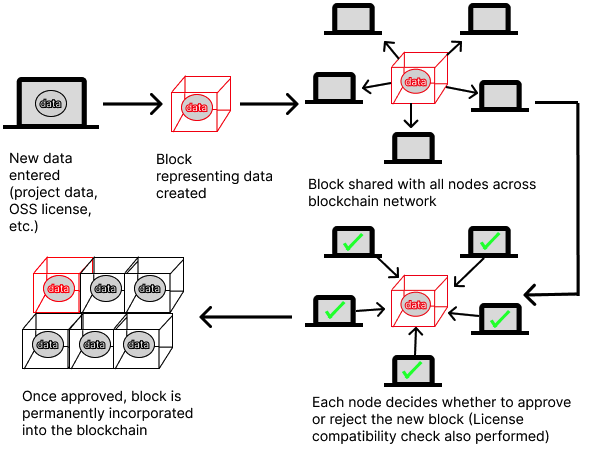}
\caption{Main workflow in \emph{FOSS-chain}.} 
\label{fig:blockchain}
\end{figure}

The core architecture of \emph{FOSS-chain} utilizes smart contracts, storage of licensing information and relevant compatibilities, and function-level hashing to guarantee license compliance. The blockchain ensures the immutability of license acceptance records and function hash logs (further explained later in the section), providing a transparent and verifiable history of each user’s interactions with the platform. \figurename~\ref{fig:blockchain} illustrates the basic workflow of \emph{FOSS-chain}. New data, such as software license agreement, are recorded as a new block. This block is then broadcast to all nodes in the decentralized network for validation. If the nodes reach consensus, the block is approved and permanently added to the chain.

Smart contracts in blockchain can store and preserve the data of OSS downloads, including the licensing terms and the function-level structure of the software project that was downloaded. Since users need to use the platform every time they want to make available a new OSS project that may have been created from scratch or may be based on existing projects, they are also required to go via the license compliance process of the platform to ensure they are respecting the licensing terms of the software they are modifying. \emph{FOSS-chain} is envisioned as a platform that manages derivative works of more than one software projects, so it is suitable for handling the OSS projects of an organization or even of the OSS community as a whole, although scaling issues arise in that case. 

\begin{figure}
\centering
\includegraphics[scale=0.29]{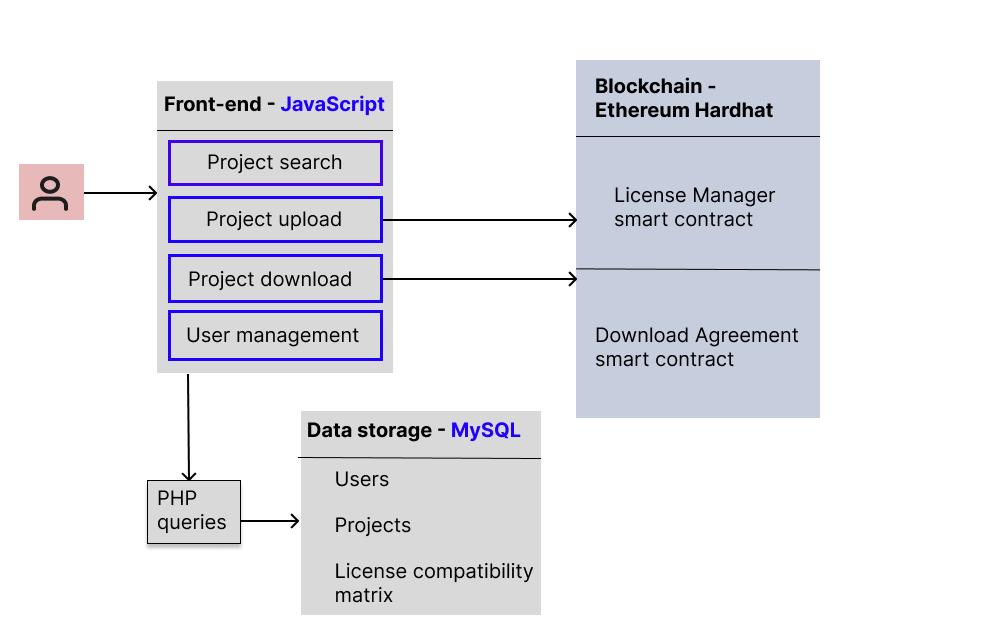}
\caption{\emph{FOSS-chain} System architecture.} 
\label{fig:architecture}
\end{figure}

The platform is accompanied with a web application that allows registered users to interact with software projects.~\figurename~\ref{fig:architecture} depicts the platform architecture with the main interacting components. Once users sign in, they are able to search for software projects, upload a new project or download an existing software project via the front-end UI (User Interface). After the user downloads the project, the front-end initiates a blockchain transaction to record the license agreement. Specifically, a smart contract agreement is triggered via the \texttt{DownloadAgreement} contract to ensure license acceptance is recorded immutably on-chain. This contract states the license agreement between the author and the downloader, who acts as licensee, and creates all function hashes of the software downloaded. These hashes uniquely identify the code at the function level. This allows the management system to identify during future uploads whether such functions were previously used.

The upload process of the system contains a verification. When a new project is uploaded, function-level hashes are extracted and are communicated to the back-end for project management and license compatibility checks. \emph{FOSS-chain} queries the blockchain to see if there is a match with any function hashes of all previously downloaded projects by the user. If the system sees equivalence, it obtains the license linked to the original function and runs a compatibility check against the license declared for the new upload. If the licenses are compatible, the project is successfully uploaded on the platform. The upload will stop if there is a license compliance issue; for instance, if a user tries to relicense a GPL code under a non-compatible license, then the user will be notified. Simultaneously, a communication with the \texttt{LicenseManager} smart contract on the blockchain is performed in order to verify and store license information. 

\subsection{Supported OSS licenses and compatibility enforcement}

\emph{FOSS-chain} currently supports 14 OSS licenses, counting also different license versions. These licenses were selected based on their popularity in OSS systems. They are listed in Table~\ref{tab:licenses-supported}, whereas their popularity according to the Open Source Initiative top licenses for 2024\footnote{\url{https://opensource.org/blog/top-open-source-licenses-in-2024}} is also indicated. In GitHub\footnote{\url{https://www.mend.io/blog/open-source-licenses-trends-and-predictions/}}, Apache-2.0 was found in 30\% of projects in 2021 and MIT in 26\%. All licenses are OSI-approved. For compatibility purposes, we have used a license compatibility matrix with the supported licenses, where we indicate for each supported license, the licenses it is compatible with. This matrix is based on a license graph for license compatibility indication from a prior work~\cite{kapitsaki2015insight}. More licenses can be added to \emph{FOSS-chain}, if license compatibility information is also available for those licenses. 

\begin{table}[!th]
\vspace{0.25cm}
\caption{Supported licenses in \emph{\emph{FOSS-chain}}.}
\label{tab:licenses-supported}
\centering
    \begin{tabular}{p{6.3cm}p{2.6cm}p{2.0cm}p{0.9cm}}
    \toprule
    \textbf{License name} & \textbf{SPDX Abbreviation} & \textbf{Category} & \textbf{Rank (OSI)} \\
    \midrule
    MIT License & MIT & Permissive & 1\\
    BSD 2-Clause "Simplified" License &BSD-2-Clause & Permissive &4\\
    BSD 3-Clause "New" or "Revised" License &BSD-3-Clause & Permissive &2\\
    Apache License 2.0 &Apache-2.0 &Permissive &3\\
    GNU General Public License v2.0 only &GPL-2.0 &Strong-copyl. &5\\
    GNU General Public License v2.0 or later &GPL-2.0-or-later &Strong-copyl.&5\\
    GNU General Public License v3.0 only &GPL-3.0 &Strong-copyl.&6\\
    GNU General Public License v3.0 or later	&GPL-3.0-or-later&Strong-copyl.& 6\\
    GNU Lesser General Public License v2.1 only &LGPL-2.1 &Weak-copyl.&8\\
    GNU Lesser General Public License v3.0 only & LGPL-3.0&Weak-copyl.&9\\ 
    Mozilla Public License 1.1 &MPL-1.1 &Weak-copyl.& --\\
    Mozilla Public License 2.0 & MPL-2.0 &Weak-copyl.&11\\
    Affero General Public License v1.0 or later	& AGPL-1.0-or-later &Strong-copyl.& --\\
    GNU Affero General Public License v3.0 & AGPL-3.0&Strong-copyl. &14\\
\bottomrule
\end{tabular}
\vspace{-0.2cm}
\end{table}

\emph{FOSS-chain} is using function-level hashing that generates a distinct hash for every function in the original software systems available in the platform, by applying the SHA-256 algorithm. As aforementioned, hashes are generated for all functions. We have created relevant regular expressions for the three supported languages of the platform: C, Java and Python. In order to detect functions in the software projects, we are using respective regular expressions for each programming language. We are providing as example the regular expression used for the C language (in PHP), while the remaining expressions are available on the GitHub repository of \emph{FOSS-chain}~\cite{replication}: 

\begin{lstlisting}[
    basicstyle=\small, %or \tiny or \footnotesize etc.
] 
/(?:function|void|int|char|float|double)\s+(\w+)\s*\([^)]*\)\s
*\{([\s\S]*?)\}/g
\end{lstlisting}

This allows the system to detect partial reuse in later uploads of the same software project. If there is a match between one or more hashes of the original software downloaded and a new software project uploaded on the \emph{FOSS-chain} platform, the back-end detects that the new project is a derivative work of a previous download. This triggers a license compatibility check after retrieving the license of the original software. \emph{FOSS-chain} examines then the compatibility matrix to assess whether the license the user intends to apply on the work is compatible with the license of the original software. This process is repeated for all projects with a match. The upload of the project is permitted on \emph{FOSS-chain} only if the licenses are compatible, or if the same license of the original software is used. Otherwise, the system prevents the upload and informs the user about the license conflict. For instance, a piece of code that is licensed under GPL cannot be relicensed under a permissive license, such as MIT, because this would violate the GPL’s copyleft requirements.

\subsection{Management of smart contracts}

As aforementioned, two main smart contracts have been introduced and used for the license compliance enforcement in \emph{FOSS-chain}: \texttt{DownloadAgreement} and \texttt{LicenseManager} contract. These smart contracts are in charge of storing and managing the licensing agreements, recording the function-level hashes of the downloaded and uploaded projects, and tracking the metadata of these projects.

\texttt{DownloadAgreement} is the first contract, which records the acceptance of the software license when the user downloads an existing software project from the platform. When someone downloads the software, this agreement stores the wallet address of the downloader, the unique identifier of the software project in the platform, the name of the license, and the timestamp of the download. These data are stored on-chain, so that they can be retrieved when required in the future for license verification purposes. The record created serves as an immutable indicator of user consent to the specific licensing terms.

The second smart contract, \texttt{LicenseManager}, manages the software project uploads, tracks the hashes of functions, and enforces license compatibility. When a user uploads a new software project, this contract will register the metadata of that project including the address of the uploader, the identifier of the project, any parent (i.e. original) project(s) it may be a derivative of, and the license selected by the uploader. More importantly, the contract stores an array of function-level hashes from the uploading project. The project’s functional logic is signified by these hashes helping with the precise detection of code reuse on the platform. \texttt{LicenseManager} is responsible for enforcing the license verification process. It is not mandatory for the user to indicate one or more parent projects when performing an upload, as the uploaded projects are checked against all projects. 

In the current version of the \emph{FOSS-chain} prototype implementation, the wallet addresses are managed manually. Specifically, the wallet addresses of newly registered users are manually entered in a configuration file by the system administrator. Platforms using blockchain usually provide automatic wallet creation and integration of wallets~\cite{biernacki2021comparative}. Future versions of \emph{FOSS-chain} will automate the generation and management of wallets, in order to increase system usability and allow scalability (e.g. for large organizations or popular OSS projects) without administrative actions.

\subsection{Implementation tools and use demonstration}

For blockchain purposes, Ethereum\footnote{\url{https://ethereum.org/}} was used, selected due to its wide popularity and adoption. We have employed the Ethereum Virtual Machine (EVM) that allows the execution of smart contracts. We have also used Hardhat\footnote{\url{https://hardhat.org/}} that is a local development environment of Ethereum without needing to use the actual network of Ethereum. Concerning the smart contracts of \emph{FOSS-chain}, they are written in Solidity which is the main programming language for contracts on Ethereum. As depicted in~\figurename~\ref{fig:architecture}, JavaScript was used for the front-end (e.g. EtherJS library), along with PHP for the back-end, while relevant data are stored on a MySQL database: user profiles, project metadata, blockchain transaction reference, and the license compatibility matrix. 

\begin{figure}
\centering
\includegraphics[scale=0.11]{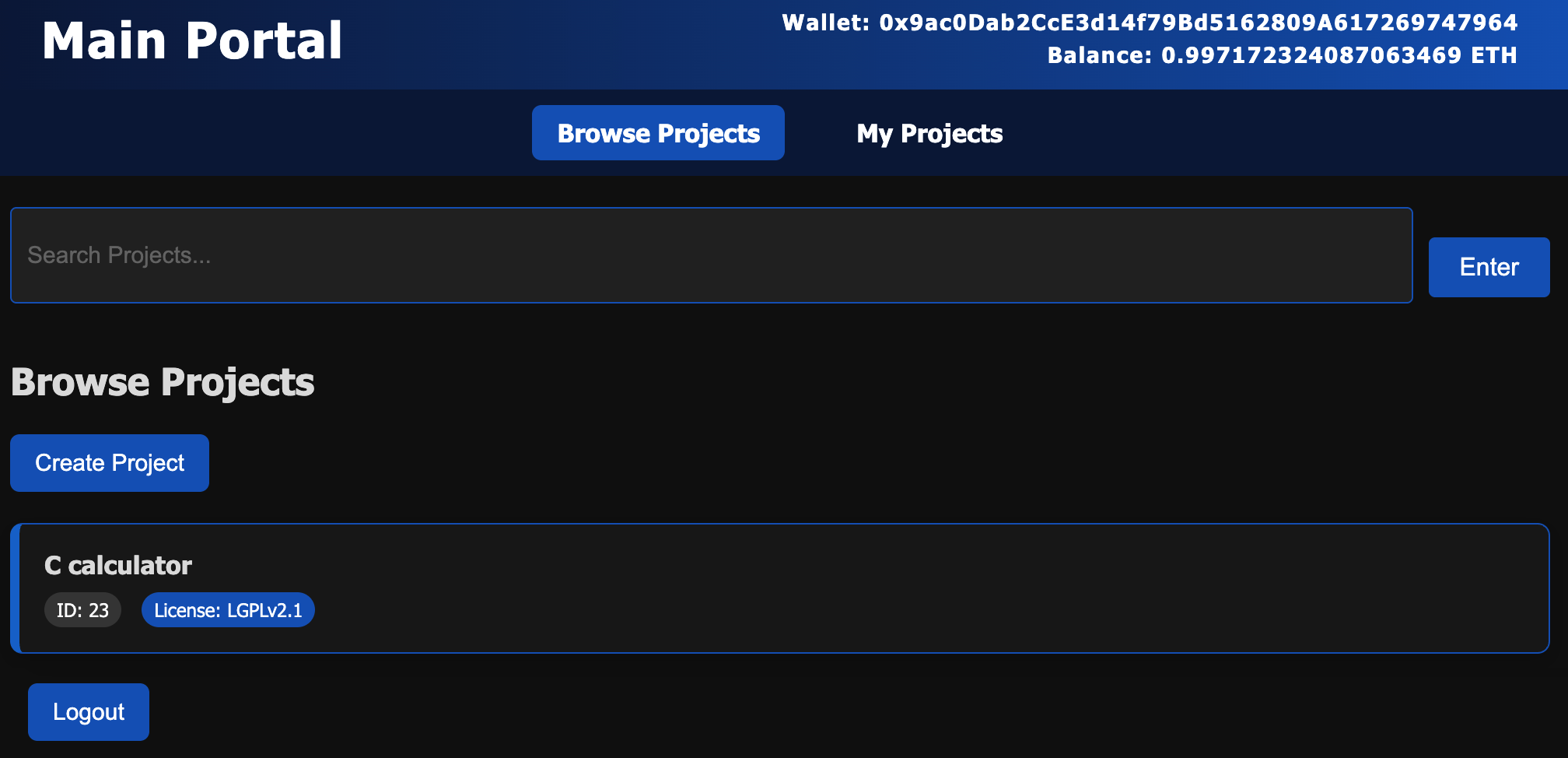}
\caption{Browsing of existing projects in \emph{FOSS-chain}.} 
\label{fig:project-browse}
\end{figure}

\begin{figure}[h!]
\centering
\begin{subfigure}{0.495\textwidth}
    \includegraphics[width=\textwidth]{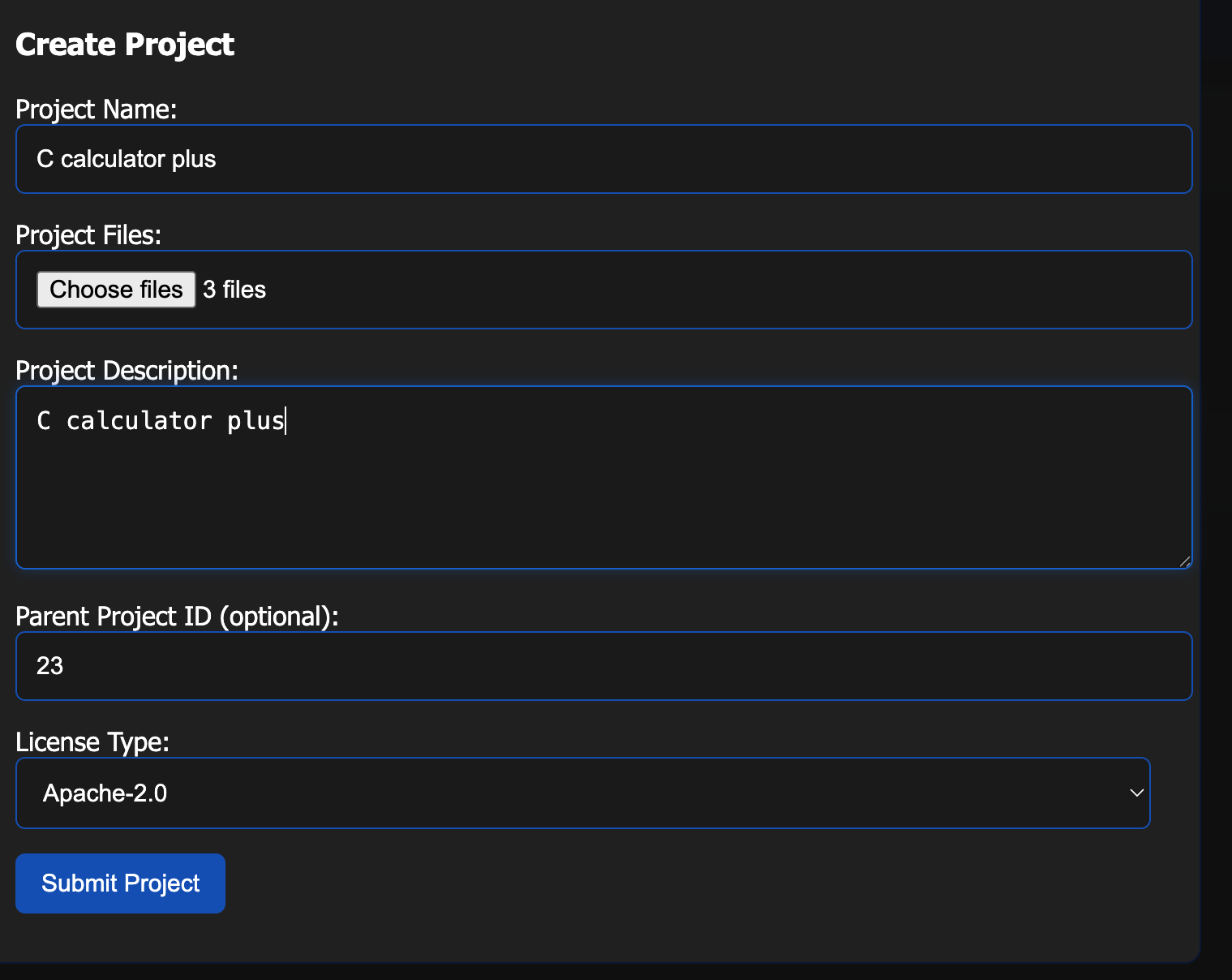}
    \caption{Upload of a derivative project.}
    \label{fig:upload-derivative}
\end{subfigure}
\begin{subfigure}{0.495\textwidth}
    \includegraphics[width=\textwidth]{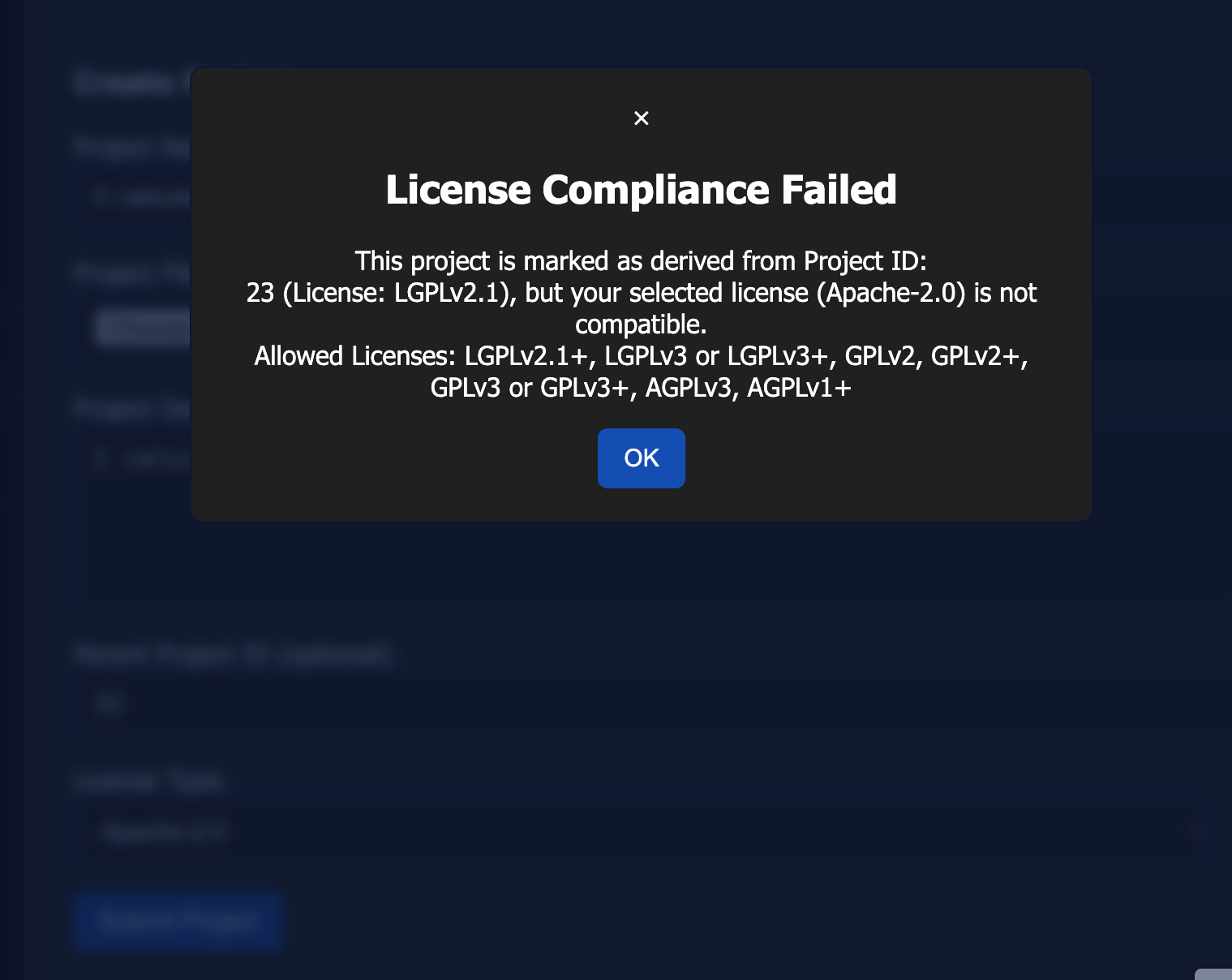}
    \caption{License conflict notification.}
    \label{fig:license-conflict}
\end{subfigure}
\caption{Upload of derivative with conflicting license in \emph{FOSS-chain}.}
\label{fig:license-check-example}
\end{figure}

A demonstration of use of the web platform of \emph{FOSS-chain} is depicted in~\figurename~\ref{fig:project-browse},~\figurename~\ref{fig:upload-derivative} and~\figurename~\ref{fig:license-conflict}. A user downloads an existing software project licensed under LGPL-2.1 from the platform after performing a relevant search (\figurename~\ref{fig:project-browse}) and uploads an updated version of the project (\figurename~\ref{fig:upload-derivative}) that causes a license violation, as the source code has been modified and the user attempts to license the LGPL-licensed original software under the Apache-2.0 permissive license. \emph{FOSS-chain} displays a notification to inform the user and does not allow the project upload on-chain (\figurename~\ref{fig:license-conflict}). \emph{FOSS-chain} also informs the user about the compatible licenses that can ne used instead. Note that the \emph{create project} form (\figurename~\ref{fig:upload-derivative}) is the same regardless of whether it concerns a completely new upload or a derivative work.

\section{User Evaluation}

\subsection{Study design}

We have performed a preliminary evaluation of the feasibility of using blockchain for OSS license compliance and the \emph{FOSS-chain} platform with a small scale user evaluation. The questionnaire created for this purpose consists of two main parts: the first part aims to collect users' general perceptions about open source software, licensing, and blockchain technology, while the second part aims to gather technical users' feedback on \emph{FOSS-chain}. Every participant regardless of their technical expertise was first asked to answer a number of general questions that touched on their familiarity and experience with OSS, their opinion on licensing and compliance, and the general concept of \emph{FOSS-chain}. After these more general sections, participants were asked whether they had a software development background. If they gave a positive answer, they were shown a short demonstration video of \emph{FOSS-chain} and after watching this demonstration, these technical participants were presented with a second set of questions aimed at evaluating the tool's clarity and practical potential. We opted for a demonstration video since \emph{FOSS-chain} has a large number of dependencies that would make the installation of the platform time consuming, which would be a restriction especially for participants with limited time. They were also invited to share thoughts on limitations, supported languages, and possible contexts for deployment. 

The participants were recruited among personal contacts of the authors via e-mail communication that targeted students and researchers within the University of Cyprus but also software engineers in the software industry in Cyprus, Greece, Germany and Sweden. No personal data were requested from users adhering to ethical standards, while the users were informed that the responses would be used solely for research purposes and for improving \emph{FOSS-chain}. In order to complete the questionnaire, the potential participants gave their consent to the above. The questionnaire consisting of 24 questions is available via Google Forms,\footnote{\url{https://forms.gle/oVHfugsJy9CbxUaH8}} and its main sections are shown in Table~\ref{tab:questionnaire}. Most closed form questions are multiple choice or in the 5-Likert scale, while the questionnaire includes also a number of open ended questions. The demonstration video of \emph{FOSS-chain} is also available on YouTube.\footnote{\url{https://www.youtube.com/watch?v=mcb1ZCnysN8}}

\begin{table}[!th]
\vspace{0.25cm}
\caption{Sections of \emph{FOSS-chain} evaluation questionnaire.}
\label{tab:questionnaire}
\centering
    \begin{tabular}{lp{3.7cm}p{1.3cm}p{6.4cm}}
    \toprule
    & \textbf{Section focus} & \textbf{\# questions} & \textbf{Example(s)}\\
    \midrule
    1.& Participants background and OSS use & 5 & What is your technical level of experience? (\emph{multiple choice})\\
    2.& Software licensing \& compliance & 4 & Do you have experience with Open Source Software licenses? (\emph{multiple choice})\\
    3. & Blockchain understanding & 2 & How well do you understand how blockchain works? (\emph{multiple choice})\\
    4. & \emph{FOSS-chain} implementation potential & 5 & If blockchain licensing was widely used, do you think it would increase compliance with Open Source Software rules? (\emph{multiple choice})\\
    5. & \emph{FOSS-chain} feedback & 8 & Did you find that the tool is easy to use? (\emph{5-Likert scale})\\
    &&&In which contexts, do you think the tool could be used? (\emph{checkboxes})\\
\bottomrule
\end{tabular}
\vspace{-0.2cm}
\end{table}

\subsection{Study results}

\begin{figure}
\centering
\includegraphics[scale=0.33]{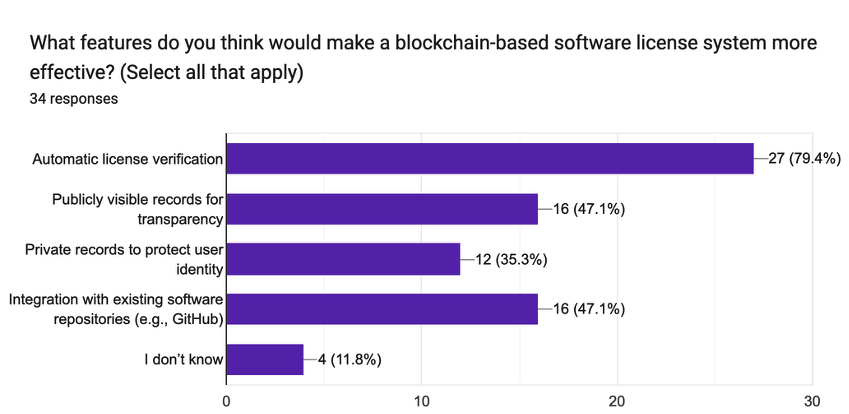}
\caption{Useful properties of a blockchain-based software license system (survey results).} 
\label{fig:blockchain-license-system-effect}
\end{figure}

A total of 34 individuals responded to the questionnaire, with most using free software frequently or every day (85.3\%) but only 32.4\% having direct experience with OSS licenses (44.1\% had limited knowledge and the remaining none). Concerning the biggest challenges in software license compliance, all participants (100\%) mentioned that `\emph{people don’t read or understand licenses}’, which was one of the options provided. Half of the participants (50\%) indicated also the absence of enforcement mechanisms as a reason. 52.9\% replied it is because people do not think licenses matter. Most survey participants do not have a good understanding of blockchain: only 5.9\% understand very well how it works and the remaining do not (64.7\%) or have a limited understanding (29.4\%). When presented with the potential of \emph{FOSS-chain} (but before watching the respective demonstration video), most participants agreed that integrating blockchain into OSS licensing can assist in preventing violations and ensuring transparency: 67.6\% said it is a good idea, while 20.6\% were not sure (the remaining 11.8\% do not find it necessary). We also presented to users a number of features that would make a blockchain-based software license system more effective and the results are depicted in~\figurename~\ref{fig:blockchain-license-system-effect}, with most participants referring to the automation of license compliance.

In terms of technical roles, the participants pool included: 8 (23.5\%) Computer Science bachelor students, 10 (29.4\%) bachelor students from non-Computer Science disciplines, 7 (20.6\%) researchers or academics and 6 (17.6\%) junior software developers. Other participants identified as senior software engineers, or data analysts (3 participants). 

\begin{figure}
\centering
\includegraphics[scale=0.38]{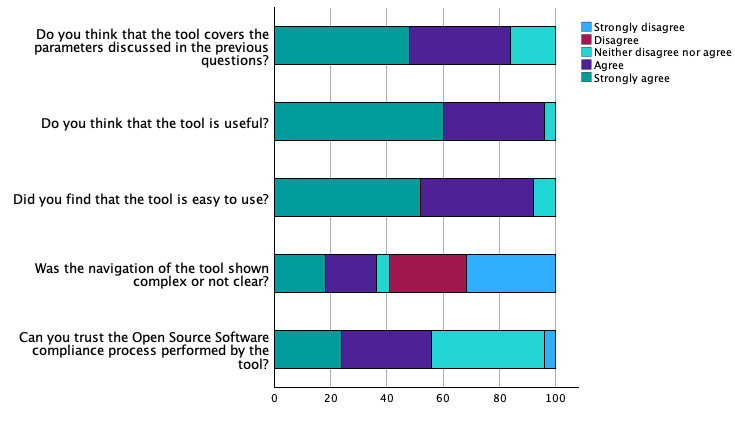}
\caption{\emph{FOSS-chain} technical users feedback (results).} 
\label{fig:main-feedback}
\end{figure}

\begin{figure}
\centering
\includegraphics[scale=0.32]{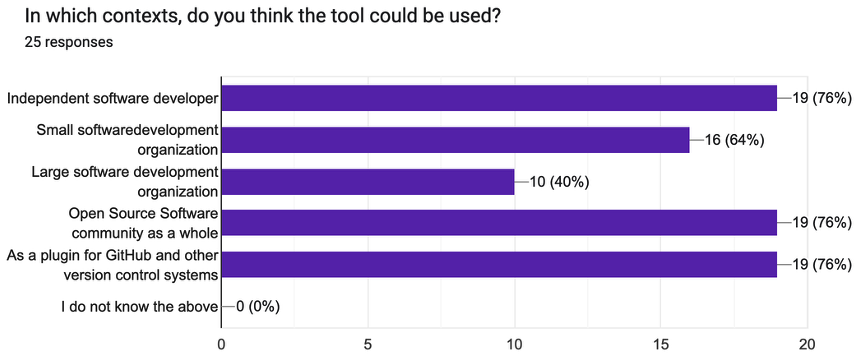}
\caption{Appropriate context of use for \emph{FOSS-chain} (survey results).} 
\label{fig:FOSS-chain-context}
\end{figure}

Concerning the results on the usage of \emph{FOSS-chain}, we analyzed separately the responses coming from technical users, so non-technical users' responses were excluded from this part of the analysis. 73.5\% indicated that they have a software development background, while the remaining 26.5\% did not, so the subsequent analysis relies on those 25 participants. We asked technical participants questions on the ease of use and complexity of \emph{FOSS-chain}, using 5 questions in the 5-Likert scale, with the results shown in~\figurename~\ref{fig:main-feedback}. The main area of improvement can be found in the navigation of \emph{FOSS-chain}, that some participants found complex. Moreover, more actions are needed to ensure users can trust the provided results. We also asked participants in which contexts they would see the platform being used, with participants results depicted in~\figurename~\ref{fig:FOSS-chain-context}. According to the participants, \emph{FOSS-chain} is more suitable for independent developers and small organizations. We ran a number of statistical tests (Kruskal-Wallis H rank-based non-parametric test) to examine whether the participants' background (e.g. their technical role or their experience with OSS licenses) affects their experience with \emph{FOSS-chain} but no statistically significant differences were found.

Technical background participants also indicated a large number of programming languages they would like to see integrated in \emph{FOSS-chain}, including Golang, JavaScript, PHP, C++, C\#, R and TypeScript. We finally gathered technical participants' view on how to improve \emph{FOSS-chain}. Seven participants provided such comments. After performing a simple qualitative analysis, the following are the main useful suggestions for future enhancements: 1) Integrate the FOSS-chain functionality into an IDE, or software editing tool. 2) Add more information on each license indicated in FOSS-chain (e.g. via URL to official license website). 3) Explain to the user why an uploaded project has been recognized as a derivative of an existing project on blockchain. This would provide the user the possibility to ask for a review of this decision.

\section{Discussion}

\textbf{Main findings.} The prototype implementation of \emph{FOSS-chain} has shown the preliminary usefulness of using blockchain for OSS license compliance purposes. The management process might be long, as the network of nodes expands, so the system might be more applicable organizational-wide and not as a solution for the OSS community as a whole, considering that in this case transaction processing will become very expensive. This is also an outcome of the results of the preliminary evaluation, as most technical background participants find \emph{FOSS-chain} more suitable for independent developers and small organizations. Thus, it might be more meaningful to apply the platform in the framework of small organizations. Its use in a large organization needs to be tested, in order to examine the scalability of the approach, whereas the same needs to be performed across organizations. Although gas pricing of Ethereum did not obstruct testing in the prototype implementation, it does pose a barrier to scaling the system, as investigated in prior work~\cite{pacheco2023makes}. While the architecture of \emph{FOSS-chain} is tailored to OSS projects, it could be expanded to proprietary software compliance, as suggested in a prior work in the framework of software piracy~\cite{shamalka2024blockchain}.

\textbf{Threats to validity.} The current implementation is limited to C, Java and Python languages, but support for additional programming languages like JavaScript, C++ and C\# can be added. This might have affected \emph{external validity} referring to the extent we can generalize our findings. A small number of developers participated in the user evaluation that was performed mainly via a video demonstration of the platform and in the framework of the University community, even though developers from the industry were also reached. We should validate our findings with an extended community of software engineers, as the current evaluation entails threats to \emph{conclusion validity}. In terms of \emph{construct validity}, the accuracy of function-level hashing as a tracker of code reuse is a limitation of the approach. While this hashing mechanism is useful to identify identical functions from other projects, it cannot detect cases where the code is altered very slightly but results to the same functionality. Therefore, the current implementation of FOSS-chain might not be able to prevent all cases of intentional license breaches. Future improvements could examine machine learning to detect code similarity. In some cases, slight modifications may not constitute a derivative work and do not require a license compatibility check, so these cases also need to be detected (e.g. if a developer changes only some variable names, or if there is only accidental equivalence of function hashes). 

A further limitation is the manual processing of wallet addresses. A system administrator must register the wallet of any user in the configuration file for the user to be able to use the platform. Although this works in a development setting, it is potentially dangerous if mishandled. Future versions must have automated wallets and decentralized identities for improved usability and reduced management costs. At the current implementation, \emph{FOSS-chain} uses a central database that will be replaced with a distributed file system in future versions.

\section{Conclusions}

In this work, we have presented \emph{FOSS-chain}, a blockchain-based solution for handling OSS license compliance through smart contracts, function-level code analysis and automatic license compliance checks. The users can upload and download a software project, while license compatibility checks are triggered whenever a software project is uploaded that is a derivative work of existing software projects on the platform. The initial user evaluation shows the usefulness of the approach and reveals areas of improvement for future work. Future work will implement the feedback gathered via the user evaluation and will focus on a more precise source code-level comparison integrating abstract syntax tree (AST) analysis or machine learning-based code similarity detection. Support for more programming languages and OSS licenses will also be added, while the consideration of multi-licensing schemes will also be examined. 

\bibliographystyle{splncs04}
\bibliography{sample-base}

\end{document}